\documentstyle[12pt]{article}
   \newcommand{\be}{\begin{eqnarray}}
   \newcommand{\ee}{\end{eqnarray}}

\begin{document}

\begin{center}
\Large{\bf  Theoretical Physics Institute\\
University of Minnesota}
\end{center}

\bigskip
\begin{flushright}
TPI-MINN-94/24-T\\
UMN-TH-1262-94\\
July 1994
\end{flushright}

\bigskip
\begin{center}
\Large {\bf Fractons in Twisted Multiflavor Schwinger Model}
\end{center}
\vskip 12pt
\begin{center} {\Large
M.A. Shifman  and A.V. Smilga\footnote{On leave of absence from
ITEP, B. Cheremushkinskaya 25, Moscow,
117259, Russia}

\vspace{.4cm}

{\normalsize {\it  Theoretical Physics Institute, Univ. of Minnesota,
Minneapolis, MN 55455}}

\vspace*{.4cm}
}

{\Large{\bf Abstract}}
\end{center}

\vskip 12pt

We consider two-dimensional $QED$ with several fermion flavors on
a finite
spatial circle. We discuss a modified version of the model with {\em
flavor-dependent} boundary conditions $\psi_p(L) = e^{2\pi ip/ N}
\psi_p(0
)$, $p = 1, \ldots , N$ where $N $ is the number of flavors. In this
case the Euclidean path integral
acquires the contribution from the gauge field configurations with
fractional topological charge being an integer multiple of $1/N$. The
configuration with $\nu = 1/N$ is responsible for the formation of
the
fermion condensate $\langle\bar{\psi}_p \psi_p\rangle_0$.
The condensate dies
out as a
power of $L^{-1}$ when the length $L$ of the spatial box is sent to
infinity.
Implications of this result for non-abelian gauge field theories are
discussed.

\newpage

\section{Motivation.}

Since the pioneering work \cite{BPST}, it is known that the Euclidean
path integrals in the gauge field
theories get contributions from  sectors with nonzero topological
charge. In the nonabelian 4-dimensional gauge theories the
topological
charge
coincides with the so called Pontryagin class and is given by the
integral
\begin{eqnarray}
\label{nu4}
\nu_4 = \frac {g^2}{32\pi^2} \int d^4x G_{\mu
\nu}^a\tilde{G}_{\mu \nu}^a .
\end{eqnarray}

It is generally assumed that the field densities $G_{\mu \nu}$
contributing
to the path integral are not singular and fall off fast enough when
$x^2 = \tau^2 + {\bf x}^2$ is sent to infinity. In this case, the
topological charge (\ref{nu4}) must be an integer which describes the
mapping $S^3 \rightarrow$ gauge group.

However, since the beginning of the eighties, different indications
have been
cropping up that the restriction $\nu_4 =$ integer is too rigid, and in
some
cases  configurations with the fractional topological charge may be
relevant.

The most explicit indication  has been unraveled while
studying
supersymmetric (SUSY) gauge theories. The simplest version of such
theory
(called
supersymmetric Yang-Mills, SYM) has the lagrangian
  \begin{eqnarray}
  \label{LSYM}
{\cal L} = - \frac{1}{4} G_{\mu \nu}^a G_{\mu \nu}^a+ \frac{i}{2}
\bar{\lambda}
{\not\!\!{\cal D}} \lambda ,
  \end{eqnarray}
where the adjoint Majorana massless field $\lambda_\alpha$ is the
superpartner of the gluon field and is called gluino. Using
supersymmetric
Ward identities one can show \cite{SYM} that, say, in the
SU(2) SUSY Yang-Mills theory the Euclidean correlator
  \begin{eqnarray}
  \label{Cdef}
C(x) = \langle\bar{\lambda}_L\lambda_R (x)
{}~~\bar{\lambda}_L\lambda_R (0)\rangle ,\,\,\, \lambda_{L,R}
=\frac{1}{2}(1\pm \gamma_5)\lambda ,
  \end{eqnarray}
does not depend on $x$.
On the other hand, it can be explicitly evaluated
at small $x$ using the instanton calculus. As a result, one gets
  \begin{eqnarray}
  \label{Cres}
C(x) = A \Lambda^6_{SYM}
  \end{eqnarray}
at all $x$; here $A$ is a  calculable number \cite{SYM},
$\Lambda_{SYM}$ is a scale parameter of SYM. The
cluster decomposition then implies that the gluino condensate is
generated,
  \begin{eqnarray}
  \label{cond4}
<\bar{\lambda}\lambda> = \pm A^{1/2} \Lambda^3_{SYM}
  \end{eqnarray}
(in Eq. (\ref{cond4}) we assumed the vacuum
angle
$\theta$ to be zero; for  a more detailed discussion of the issue of the
$\theta$
dependence and the double-valuedness of
$\langle \bar\lambda\lambda\rangle$ see below).

And here we run into a paradox. The point  is that we cannot
generate a non-vanishing
gluino condensate
  \begin{eqnarray}
  \label{cvar}
<\bar{\lambda}\lambda> = - \frac \partial {\partial m} \ln Z_m
|_{m=0}
  \end{eqnarray}
($Z_m$ is the partition function of the theory involving a small
Majorana
mass term which is treated as a source) directly from the path
integral
if we integrate only over the field
configurations with the integer topological charge. Due to the index
theorem \cite{index},
the configuration with a given $\nu_4$ involves $N_c$ pairs of gluino
zero modes (this number is just the Dynkin index of the adjoint
representation)
which it ``too much". The corresponding contribution to the partition
function involves a suppression factor
  \begin{eqnarray}
  \label{Zm}
Z_m(\nu_4) \propto m^{|\nu_4|N_c}
  \end{eqnarray}
and the contribution of all nonzero integer $\nu_4$ to the fermion
condensate (\ref{cvar}) is suppressed.

To be more precise, one should regularize the theory by putting it in
a
large but finite Euclidean volume $V \gg \Lambda_{SYM}^{-4}$ and
take the
limit $m \rightarrow 0$ while keeping the volume fixed. Due to the
{\em
absence of massless states} in the physical spectrum of
the supersymmetric Yang-Mills
theory,
the chiral limit is smooth here, see Refs. \cite{LS,QCD2} for a detailed
discussion. As a matter of fact, one can show \cite{S} that in SYM the
value of the gluino condensate does not depend on the volume at all.
When considered in a small three-dimensional volume, $V_3<
\Lambda_{SYM}^{-3}$, the system becomes quasiclassical, and the
path integral {\em must} be saturated by saddle points. Then the fact
that $\langle\bar\lambda\lambda\rangle\neq 0$ proves that
the field configurations with fractional topological charge must
contribute to the path integral.

The partition function in the topologically trivial sector is expanded
in
even powers of $m$. Indeed, the spectrum of the Euclidean Dirac
operator on  a
topologically trivial background involves only nonzero eigenvalues
which
come in pairs ($\mu_n, -\mu_n$) so that the determinant of the
Dirac
operator $\propto \prod_n (\mu_n^2 + m^2)$ ~ does not produce a
linear in
$m$ term in the expansion. Hence, it also does not contribute to the
condensate.

The only way out is to admit the existence of  field configurations
with a
fractional topological charge which contribute to the path integral. If
$\nu_4 = \pm 1/N_c$, $Z_m(\nu_4) \propto m $, i.e.  just what is
required
to generate the fermion condensate (\ref{cvar}).

But what are these configurations? For the $SYM$ theory with the
{\em
unitary}
gauge group, one can give at least a partial answer to this question.
As
has been observed by 't Hooft long ago \cite{Hooft}, such
configurations
arise, indeed, if we define the theory not on a 4-dimensional large
sphere
and not on a cylinder $S^3 \otimes R^1$ (the theorem that $\nu_4$
should be
integer applies only to these cases) but on a 4-dimensional
torus,
and allow for non-standard (twisted) boundary conditions for the
gauge
field, the so called torons. Witten further studied the twisted
construction on $T^3 \otimes R^1$ \cite{witten} and revealed the
field configurations between which the torons interpolate. Cohen and
Gomez showed that
torons do indeed  contribute to  the gluino condensate \cite{Cohen}.
In Ref. \cite{LS}
the volume  and
mass dependence of the partition function in the toron sector has
been
determined. Zhitnitsky \cite{Zhit} suggested another type of the field
configurations with the fractional topological charge. They are
defined on
$S^4$ but are singular (topological classification is valid only for
smooth fields). The singularity is not too strong, however, so that the
action of such singular fracton is still finite.

The existence of all these solutions crucially depends, however, on
the
presence of the center subgroup in the gauge group. For $SU(N)$ the
center is $Z_N$. It is even better to say that the proper gauge group
of
the theory is not $SU(N)$ but rather $SU(N)/Z_N$ (the elements of
the
center do {\em not} transform the fields in the adjoint color
representation).
The factorization over $Z_N$ makes $\pi_1$({\it gauge group})
nontrivial
which brings about new topological possibilities \cite{bubble,QCD2}.

The real problem appears when we consider the SYM theories with
orthogonal
gauge groups $O(N),\,\,\,  N\geq 5$ \cite{ort,QCD2}. The Ward
identities + instanton  calculus imply a non-vanishing gluino
condensate exactly by the same token as in the unitary case.
The index theorem dictates the presence of $2(N-2)$ gluino zero
modes
in the
instanton background -- too much to generate the fermion
condensate.
On the other hand, the center is either absent here (odd $N$) or is too
small ($Z_2$ for even $N$), and we cannot construct a configuration
which
carries just 2 fermion zero modes and do not understand the
mechanism for
generating the fermion condensate. The same situation takes place
for exceptional groups \cite{morozov}.

A similar problem appears also in two-dimensional $QCD$ with
the adjoint
Majorana fermions \cite{QCD2}. The paradox arises here already for
the unitary
gauge groups if the number of colors is 3 or more. The instantons
(they
appear in the {\em adjoint} $QCD_2$ due to the presence of
nontrivial
$\pi_1[SU(N)/Z_N] = Z_N$ ) involve here $2(N_c-1)$ fermion zero
modes and
cannot generate the fermion condensate for $N_c \geq 3$. On the
other  hand,
independent arguments (based on the bosonized representation of
the theory)
do imply the presence of the fermion condensate. There are no
massless
states in the spectrum, the chiral limit $m \rightarrow 0$ is smooth,
and
the only way to generate the nonzero condensate is to admit the
existence
of some field configurations carrying just 2 zero modes and
contributing to the Euclidean path integral. Such
configuration are not known so far, however.

In this paper we discuss instead the simplest
possible model where a similar paradox can be formulated {\em and}
successfully resolved.  We hasten to add that the results reported
here literally speaking add little to this issue in the non-abelian
theories. The paradox is still there. Some parallels, however, seem to
be instructive, and may provide insights in the non-abelian case in
future.

The model to be considered below is basically the Schwinger model
(QED in $1+1$ dimensions) treated in a finite spatial box of dimension
$L$. The lagrangian
includes not one (as in the original Schwinger model) but several
fermion flavors:
  \begin{eqnarray}
  \label{LSM}
{\cal L} = - \frac 14 F_{\mu \nu}^2 + i \sum_{p=1}^{N}\bar{\psi}_p
{\not\!\!{\cal D}} \psi_p - m \sum_{p=1}^{N}\bar{\psi}_p
 \psi_p
\end{eqnarray}
where ${\cal D}_\mu = \partial_\mu - igA_\mu$, $N$ is the number
of flavors,  and the charge $g$
has  dimension of mass and defines the characteristic mass scale
of the  theory
(it plays the same role as $\Lambda_{QCD}$ and $\Lambda_{SYM}$).
The  fermion mass $m$ is chosen to be much less than $g$ and will
be  treated as a
perturbation, to be put equal to zero at the very end. The nontrivial
picture which resembles the situation in the
non-abelian gauge theories outlined above arises if we impose {\em
flavor-dependent} spatial boundary conditions on the fermion fields.
Namely,
  \begin{eqnarray}
  \label{bc}
\psi_p(x=L) = e^{2\pi ip/N} \psi_p(x=0),\,\,\, p =1,2,..., N,
  \end{eqnarray}
where $L$ is the length of the circle where the theory is defined.
This non-standard boundary condition, a heart of our construction,
will lead to a dramatic deviation from the standard description of the
Schwinger model \footnote{Twisted boundary conditions on fermions
conceptually resembling ours were discussed previously in QCD
with $N_c=N_f$ in Ref. \cite{CG}.}.

Let us remind that the theory involves instantons, and the usual
topological arguments
require
  \begin{eqnarray}
  \label{nu2}
\nu_2 = \frac{g}{4\pi} \int \epsilon_{\mu \nu} F_{\mu \nu} d^2x
  \end{eqnarray}
(the two-dimensional analog of the Pontryagin class) to be an integer.
The index theorem implies then the existence of a complex zero
mode in the
instanton background for each fermion flavor. Thus, the partition
function is
  \begin{eqnarray}
  \label{ZmSM}
Z_m(inst) \propto m^{N} ,
  \end{eqnarray}
and the fermion condensate seemingly cannot be generated.

Independent arguments presented below show, however,  that the
condensate {\em is} generated under the boundary conditions
(\ref{bc}).
This fact can be explained only if the path integral gets a
contribution from  the
field configurations with a fractional topological charge.

We will show that in the case at hand such fractons are, indeed,
present. We display
their origin and calculate explicitly the path integral on the fracton
background (since the model in the small mass limit is exactly
soluble,  all
calculations can be done analytically and the results are {\em exact}).

The main difference between  this toy model and complicated
non-abelian models
discussed above is that here the fermion condensate is  essentially a
finite-volume effect. It dies out when the length of the box is sent to
infinity. (This is, of course, natural. If the box is large, the
nonstandard boundary conditions (\ref{bc}) should not affect local
physics. For large $L$, the theory has the same properties as the
standard  multiflavor Schwinger model where no condensate
appears). Apart from this distinction other qualitative features of our
construction will hopefully serve as a prototype for the SYM
theories.

The paper is organized as  follows. In Sect. 2 we remind
in brief  the solution of  the standard multiflavor Schwinger model
(with the trivial boundary conditions). In Sect. 3 we start our
discussion of  the twisted model and consider first the case of a small
spatial circle,  $gL
\ll 1$, where the situation is especially simple and transparent. In
Sect. 4 we calculate the fermion condensate at arbitrary $L$ via the
correlator
$ \langle\bar{\psi}\psi (\tau) ~~\bar{\psi}\psi (0)\rangle$  at large
Euclidean
time
$\tau$. In Sect. 5 we rederive this result defining and calculating the
corresponding path integral in the fracton background. General
discussion is given in Sect. 6.

\section{Multiflavor Schwinger model}

\setcounter{equation}0

Consider first the theory with the lagrangian (\ref{LSM}) in the
two-dimensional
space-time without boundaries. In the massless limit, the tree
lagrangian
has the symmetry $SU_L(N) \otimes SU_R(f) \otimes U(1)
\otimes U(1)$. As well-known, the
$U_A(1)$ part of it is anomalous,
  \begin{eqnarray}
  \label{anom}
\partial_\mu j_\mu^5 =
\frac{g}{2\pi}\epsilon_{\alpha\beta}F_{\alpha\beta} ,
  \end{eqnarray}
and is absent in the full quantum theory
(for a review see, e. g. \cite{Shif}). The $SU_A(N)$
part of the symmetry is broken explicitly by the mass term in the
lagrangian. But if the mass is small, $m \ll g$, the breaking is weak.

The spectrum of the theory involves a ``massive photon" (a
flavor-singlet scalar meson) and $N^2-1$ light mesons belonging to
the
$SU(N)$  multiplet
of the flavor group \cite{colSM}. The photon mass is
  \begin{eqnarray}
  \label{muN}
\mu^2 = \frac {Ng^2}\pi + {\cal O}(mg)
  \end{eqnarray}
while the mass of light mesons is
  \begin{eqnarray}
  \label{mlight}
m^2_{light}  \sim g^{1/(N+1)}m^{N/(N+1)} .
  \end{eqnarray}
The multiplet of light mesons superficially resembles a
pseudogoldstone multiplet  in the
standard 4-dimensional $QCD$; there is  an important distinction,
however. In the
chiral limit $m \rightarrow 0$, not only the mass of these
``pseudo-pseudogoldstones" turns to zero but, also, their coupling to
the  massive sector vanishes (in contrast to pions which {\em are}
coupled to massive hadrons). The fact that the massless states
become sterile is specific for two dimensions -- a consistent
two-dimensional field theory
with massless interacting bosons does not exist \cite{Coleman}.

The same Coleman theorem prohibits the formation of the fermion
condensate
in the massless Schwinger model with several flavors. Indeed, the
condensate would break spontaneously the symmetry $SU_L(N)
\otimes SU_R(N)
$ down to $SU_V(N)$ which would lead to appearance of non-sterile
Goldstone particles in the spectrum  not allowed in two
dimensions.

If $m \neq 0$ and the symmetry $SU_A(N)$ is already broken
explicitly,
the formation of the condensate is allowed. It has a funny
non-perturbative
dependence on the fermion mass \cite{cond},
   \begin{eqnarray}
  \label{condm}
\langle \bar{\psi} \psi\rangle \sim g^{2/(N +1)}m^{(N-1)/(N+1)} .
   \end{eqnarray}
It is instructive to consider is the correlator
$$
C_N(x) = \langle T\{\bar{\psi}_L\psi_R(x)~
\bar{\psi}_R\psi_L(0)\}\rangle_0
$$
in the massless theory.
At
large $x$, $gx \gg 1$, it falls off as a power,
  \begin{eqnarray}
  \label{cinfv}
C_N(x)  \sim g^{2/N} x^{2/N - 2} .
  \end{eqnarray}
One may ask how one can reconcile this power falloff of the
correlator which may
be only due  to massless intermediate states with the fact that the
massless
states decouple in the chiral limit. The answer is as  follows. The
Coleman theorem dictates that all matrix elements
$\langle 0|\bar{\psi}\psi |n\phi_{light}\rangle $ vanish in the chiral
limit. And they do. Let us allow,
however, for a small nonzero mass $m$. The correlator (\ref{cinfv})
is
saturated by the sum over all possible intermediate states involving
light  mesons. The smaller  the mass, the larger the characteristic
number $n$ of "pseudo-pseudogoldstones" is in the intermediate
state.
As a result, the
chiral limit of the whole sum is nonzero (see Eq. (5.6) in
Ref. \cite{cond}).

Anyway, the correlator tends to zero in the limit $x \rightarrow
\infty$
and the fermion condensate vanishes. No contradiction with the
instanton counting arises.

It is interesting, however, to derive the result (\ref{cinfv}) without
using the
bosonization approach (as was done in \cite{cond}) but directly from
the
Euclidean path integral involving fermion and gauge fields. One can
then
ask what characteristic field configurations are contributing to the
path
integral. In answering this question fractons popped out
unexpectedly \cite{fract}
\footnote{Fractons also show up in the standard one-flavor
Schwinger model in the expectation value
$\langle\bar{\psi}(x) \exp\{
ie \int_0^x A_\mu dx_\mu \} \psi(0)\rangle $.}.

The picture outlined above refers to the infinite volume. Let us
proceed now to the theory defined in a finite volume.

If the volume where the theory is defined is large the quantum
fluctuations are
strong and the quasiclassical picture does not work. The situation is
much simpler
and more clear when the theory is defined on a cylinder $S^1 \otimes
R^1$
where $R^1$ is a small spatial circle with the length $L \ll g^{-1}$.

It will be shown in Sect. 4 that in the small $gL$ limit the path
integral for the correlator
  \begin{eqnarray}
  \label{ctau}
C(\tau_0) = \langle\{\bar{\psi}_L\psi_R(\tau_0,0)~
\bar{\psi}_R\psi_L(0)\}\rangle
  \end{eqnarray}
in the multiflavor Schwinger model defined on a circle ($\tau_0 \gg
g^{-1}$
is the  Euclidean time) is saturated by the gauge field configuration
  \begin{eqnarray}
  \label{faf}
A_1(\tau) = A_1^{fr}(\tau) - A_1^{fr}(\tau - \tau_0)
  \end{eqnarray}
where
  \begin{eqnarray}
  \label{Afr}
A_1^{fr}(\tau) = \left\{ \begin{array}{c}
\frac \pi{NgL}\exp\{\mu\tau\},~~~~~~~~~\tau \leq 0 \\~~\\
\frac \pi{NgL}[2-\exp\{-\mu\tau\}],~~~~~~~~~\tau \geq 0
\end{array} \right.
  \end{eqnarray}
(the gauge $A_0  = 0$ is chosen; $\mu$ is the photon mass defined in
(\ref{muN}); note that the contribution (\ref{faf}) does not depend on
$x$
as it should be on a small circle). The quantum fluctuations around
the
stationary  quasiclassical configuration (\ref{faf}) are suppressed
with
respect to the characteristic amplitude of the solution (\ref{faf}) by
the
factor $\sim \sqrt{gL} \ll 1$ \cite{inst}.

The function (\ref{faf}) is plotted in Fig. 1. We see that the total
topological charge (\ref{nu2}) of this configuration is zero:
   \begin{eqnarray}
   \label{nufaf}
\nu = \frac{gL}{2\pi} \int_{-\infty}^\infty \dot{A}_1 d\tau ~~=~~
\frac{gL}{2\pi} \left[ A_1(\infty) - A_1(-\infty) \right] = 0 .
   \end{eqnarray}
That is what one could expect right from the beginning as the
correlator
(\ref{ctau}) gets contributions only from the topologically trivial
sector.
However, the zero result is due to cancellation of two
distinct
nonzero contributions with the opposite signs coming from different
regions in
the integral (\ref{nufaf}): $\nu = 1/N$ due to the region $\tau \sim
0$ and
$\nu = -1/N$ due to the region $\tau \sim \tau_0$
(it is assumed that $\tau_0 g\gg 1$).

If $N>1$ the individual function (\ref{Afr}) of which the full
stationary solution
(\ref{faf}) is composed presents a {\em fracton}, the configuration
with a fractional topological charge.

The configuration (\ref{faf}) is the stationary point of the path
integral both in the standard multiflavor Schwinger model (with
periodic
boundary conditions for all flavors) and in the twisted Schwinger
model (with the boundary conditions (\ref{bc})).

The behavior of the correlator (\ref{ctau}) is, however, completely
different in these two cases. When boundary conditions are standard,
the
correlator falls off exponentially at large $\tau_0$, $\tau_0 \gg L $
(if $\tau_0 < L$ the fall-off is power-like, see above),  the fermion
condensate
is zero, and, though fractons appear as an ingredient in the relevant
field
configuration (\ref{faf}), there is no need to consider an {\em
isolated}
fracton. The total topological charge of field configurations in the
path
integral is always integer.

For the twisted model, the situation is, however, different. The
correlator
(\ref{ctau}) tends to a constant at large $\tau_0$ in this case. The
cluster
decomposition property then implies the appearance of the fermion
condensate and
we {\em need} an isolated fracton to generate it.

The picture in the twisted model is  similar to what we had in
the standard Schwinger model with {\em one} flavor, see Ref.
\cite{inst}.
(In this work the Schwinger model at large temperature
was considered, on the
infinite circle. But the results can be easily translated to the theory at
zero
temperature and on a small circle after rotation in the
Euclidean space by $\pi/2$). The correlator (\ref{ctau}) in Ref.
\cite{inst} was saturated by
an  instanton-antiinstanton configuration yielding a constant at
large $\tau_0$. The fermion condensate itself was generated by an
isolated instanton.

\section{Theory on  small circle: Hamiltonian
approach}

\setcounter{equation}{0}
In this section we calculate the fermion condensate in
the model defined on a small spatial circle. The case $gL \ll 1$ is
especially simple because the higher Fourier harmonics decouple
here (with reservations to be specified later) and we are left with a
quantum-mechanical problem which is easily analyzed within the
Hamiltonian approach.
To make discussion  self-contained and transparent we first
describe the well-known situation in the standard Schwinger model
with one fermion flavor (see e.g. \cite{Shif}) and then pass to a more
complicated twisted multiflavor case.

\subsection{The standard Schwinger  model}

The hamiltonian of the theory is
  \begin{eqnarray}
  \label{Ham}
H = \int_0^L dx \left\{ \frac {E^2(x)}2 - i\bar{\psi}(x) \gamma_1
[\partial_1 - igA_1(x)]\psi(x) \right\}
  \end{eqnarray}
where $E(x) = -i\partial/\partial A_1(x)$ is the electric field. Not all
eigenstates of the hamiltonian are admissible but only those which
satisfy
the Gauss law constraint
  \begin{eqnarray}
  \label{Gauss}
G(x) = \partial_x E(x) - g\bar{\psi}(x) \gamma_0 \psi(x) = 0 .
  \end{eqnarray}
Two-dimensional $\gamma$ matrices may be chosen
(in the Minkowski space) as $\gamma_0
= \sigma_2,
 \gamma_1 = i\sigma_1$; $\gamma^5 =
\sigma_3$.
 Let us impose the periodic boundary conditions on all fields and
expand
them in the Fourier series
  \begin{eqnarray}
  \label{Fourier}
A_1(x) = \sum_{k=-\infty}^{\infty} A_1^{(k)} {\rm e}^{2\pi ikx/L} ,
\nonumber \\
\psi(x) = \sum_{k=-\infty}^{\infty} \psi^{(k)} {\rm e}^{2\pi ikx/L} .
  \end{eqnarray}

When $gL$ is small, one can separate all dynamical variables in two
classes:
the {\em slow} variables
$$
A_1^{(0)}\equiv a
$$
 (we shall shortly see that its
characteristic excitation energy is of order $g$) and the fast variables
$\psi^{(k)}$ and $A_1^{(k\neq 0)}$ with characteristic excitation
energies
$\sim 1/L \gg g$. Then we can treat the system in the spirit of the
Born-Oppenheimer approximation --
integrate over fast variables and solve the Schr\"{o}dinger
equation for the effective hamiltonian thus obtained which depends
only on
the slow variable $a$.

One could adopt a slightly different formulation of the Hamiltonian
approach (see, e. g. \cite{Shif}). Instead of imposing the Gauss law
constraints on the states one could eliminate the gauge degrees of
freedom of the photon field. Namely, all non-zero modes
$A_1^{(k\neq 0)}$ can be gauged away. Our system then would
consist of the modes of the fermion field $\psi^{(k)}$  coupled to
$a$ and interacting with each other {\em via} the Coulomb
interaction. The latter is non-local in space, instantaneous in time
and can be neglected altogether provided that $gL\ll 1$. After
integrating out the fast variables we would arrive at the same
description of the slow variable $a$.

In the lowest order in the Born-Oppenheimer expansion, the induced
potential $V^{eff}(a)$ is given just by the sum of the zero-point
energies of the fermion oscillators estimated in a constant
(time-independent) background $a$
(the higher harmonics $A_1^{k\neq 0}$ are not coupled directly to
$a$ and can be neglected in this order). The structure of fermion
levels as a function of $a$ is shown in Fig. 2.

The fermion ground state corresponds to filling out all levels with the
negative energies (the Dirac sea) and leaving empty all levels with
the
positive energy. From the picture in Fig. 2 it is clear that the
minimum
of the induced potential $V^{eff}(a)$ is achieved at
  \begin{eqnarray}
  \label{Amin}
a = \frac \pi{gL} (2n+1),~~~~~~~~n ~~{\rm integer}.
  \end{eqnarray}
As usual, the sum of the zero-point energies involves an infinite
constant
which should be subtracted. It is convenient to normalize
$V^{eff}(a)$
at its minimal value achieved in the points (\ref{Amin}). Then the
effective Born-Oppenheimer hamiltonian calculated in the vicinity of
the $n$-th minimum takes the form
  \begin{eqnarray}
  \label{Heff}
H_n^{eff}(a) = \frac{1}{2L}\left( -i d /da\right)^2 + \frac{\mu^2L}{2}
\left(
a - \frac {\pi(2n+1)}{gL} \right)^2
  \end{eqnarray}
where $\mu^2 = g^2/\pi$.
 Characteristic excitation energies of this oscillator hamiltonian are of
order $g$ as was mentioned earlier.
In the region
$$
a \sim \frac{\pi(2n+1)}{gL}
$$
the ground state
wave
function of the full hamiltonian (\ref{Ham}) is the product of the
ground
sate wave function of the slow hamiltonian (\ref{Heff}) and the fast
fermion
wave functions corresponding to filling out all negative levels. We
have
  \begin{eqnarray}
  \label{wave}
|vac_n\rangle = \left(\frac{L\mu}{\pi}\right)^{1/4}{\rm
e}^{-\frac{\mu L}2 \left[
a - \frac {\pi(2n+1)}{gL} \right]^2} \prod_k \psi_L^{(k)}
\prod_k \psi_R^{(k)} ,
  \end{eqnarray}
where
$$
\psi_{L,R} = \left( \frac{1 \pm \gamma^5}2 \right) \psi
$$
are the upper and the lower components of the spinor field $\psi$,
correspondingly. The fermion part of the wave function (\ref{wave})
is written symbolically: which particular levels have negative
energies depends on $n$; for more details see Ref. \cite{Shif}.
The freedom in choosing $n$ corresponds to the symmetry of the full
hamiltonian (\ref{Ham}):
  \begin{eqnarray}
  \label{T}
S:~~~~~~~
\left\{
\begin{array}{c}
A_1(x) \rightarrow A_1(x) + \frac {2\pi}{gL}\\~~~\\
\psi(x) \rightarrow {\rm e}^{2\pi i x/L} \psi(x)
\end{array}\right.
  \end{eqnarray}
so that $S|vac_n\rangle  = |vac_{n+1}\rangle$.
This symmetry presents the so called {\em large} gauge
transformation. The
Gauss law constraint (\ref{Gauss}) requires the invariance of the
wave
functions with
respect to infinitesimal gauge transformations but does not say
anything
about transformation properties of the wave
functions under the action of the transformation $S$. We can impose,
however, an additional restriction (called the {\em superselection
rule}) and
consider the sector of the theory with all wave functions satisfying
the
requirement
  \begin{eqnarray}
  \label{ssrule}
\Psi(A^S, \psi^S) = {\rm e}^{i\theta} \Psi(A, \psi) .
  \end{eqnarray}
Then the action of any local operator on the wave function belonging
to the
class (\ref{ssrule}) leaves it within the same class. The ground state
wave
function in the sector with the given $\theta$ has the form
   \begin{eqnarray}
   \label{vactet}
\Psi_\theta\equiv
|vac_\theta \rangle = \sum_{n=-\infty}^\infty {\rm e}^{in\theta}
|vac_n\rangle ;
   \end{eqnarray}
$\theta$ has the same meaning as in the Yang-Mills theory and is
called the
vacuum angle.

Everything is ready  now for calculating the fermion
condensate.  By convoluting the wave functions (\ref{wave}) we get
\footnote{Here $\psi$ and $\bar{\psi}$ stand for the Heisenberg field
operators defined in a usual way so that $\psi_R$ involves
annihilation
operators for the right-handed states and $\bar{\psi}_L$
creation operators for the left-handed states. The fields
$\psi$ and $\bar{\psi}$ are not to be  mixed  up with the
holomorphic field variables appearing in Eq.(\ref{wave}).}
  $$
\langle \bar{\psi}_R \psi_L\rangle_\theta =
\frac {\langle vac_\theta |\bar{\psi}_R \psi_L|vac_\theta\rangle}
{\langle vac_\theta |vac_\theta\rangle } = {\rm e}^{-i\theta}
\langle vac_n |\bar{\psi}_R \psi_L|vac_{n-1}\rangle
$$
\begin{eqnarray}
   \label{condLR}
 = {\rm e}^{-i\theta} \frac 1L \exp\{
- \frac {\pi}{\mu L} \}
  \end{eqnarray}
and, correspondingly,
  \begin{eqnarray}
   \label{condRL}
\langle \bar{\psi}_R \psi_L\rangle_\theta = {\rm e}^{i\theta} \frac
1L
\exp\{
- \frac {\pi}{\mu L} \} .
  \end{eqnarray}
The exponent in Eqs. (\ref{condLR}), (\ref{condRL}) is essentially the
action
of the instanton  -- the Euclidean field configuration
which interpolates between the adjacent minima
(\ref{Amin})  and minimizes the effective action \cite{inst}. The same
result for the fermion condensate can be obtained through
bosonization \cite{hosotani,cond}.

\subsection{Twisted model}
Let us consider first, for simplicity, the model with two fermion
flavors.
The hamiltonian of the model is
$$
H = \int_0^L dx \left\{ \frac {E^2(x)}2 \right. -
$$
 \begin{eqnarray}
  \label{Ham2}
\left.
i\bar{\psi}(x) \gamma_1
[\partial_1 - igA_1(x)]\psi(x)
- i\bar{\chi}(x) \gamma_1
[\partial_1 - igA_1(x)]\chi(x) \right\}
  \end{eqnarray}
with the constraint
  \begin{eqnarray}
  \label{Gauss2}
G(x) = \partial_x E(x) - g\bar{\psi}(x) \gamma_0 \psi(x)
- g\bar{\chi}(x) \gamma_0 \chi(x) = 0 .
  \end{eqnarray}
The twisted boundary conditions
$$
\psi(x=L) = \psi(x=0) ,
$$
\begin{eqnarray}
  \label{bc2}
\chi(x=L) = -\chi(x=0)
  \end{eqnarray}
are chosen. As earlier, we assume $gL \ll 1$  which allows us to
repeat
the same Born-Oppenheimer type analysis as in the previous
subsection.
The  structure of the fermion levels is shown in Fig. 3. We see that
the minima of the effective potential $V^{eff}(a)$ occur now at
  \begin{eqnarray}
  \label{Amin2}
a= \frac \pi{2gL} (2n+1),~~~~~~~~n ~~{\rm integer}.
  \end{eqnarray}
The distance between the adjacent minima is now only $\pi/gL$,
twice  smaller than in the standard Schwinger model (or in the
multiflavor Schwinger model with the
standard
boundary conditions). The appearance of the  new minima is due to a
new global  symmetry of the hamiltonian (\ref{Ham2}):
  \begin{eqnarray}
  \label{Ttil}
\tilde{S}:~~~~~~~\left\{
\begin{array}{c}
A_1(x) \rightarrow A_1(x) + \frac {\pi}{gL}\\~~~\\
\psi(x) \rightarrow {\rm e}^{\pi i x/L} \chi(x)\\~~~\\
\chi(x) \rightarrow {\rm e}^{\pi i x/L} \psi(x)
\end{array}\right.
  \end{eqnarray}
This new symmetry is a combination of two ``non-symmetries":
a previously forbidden gauge transformation and a flavor SU(2)
rotation $\psi
\leftrightarrow \chi$
forbidden by itself by the twisted boundary conditions.
The double application of $\tilde S$ is equivalent to $S$.
The symmetry
  \begin{eqnarray}
  \label{T2}
S = \tilde{S}^2=\left\{
\begin{array}{c}
A_1(x) \rightarrow A_1(x) + \frac {2\pi}{gL}\\~~~\\
\psi(x) \rightarrow {\rm e}^{2\pi i x/L} \psi(x)\\~~~\\
\chi(x) \rightarrow {\rm e}^{2\pi i x/L} \chi(x)
\end{array}\right.
  \end{eqnarray}
persists in the theory with any choice of boundary conditions. The
square
root $\tilde{S}$ can be extracted only under the choice (\ref{bc2}).
The
symmetry $\tilde{S}$ is {\em} not just a large gauge transformation,
unlike
$S$.

 The effective hamiltonian in the vicinity of one of the minima is
  \begin{eqnarray}
  \label{Heff2}
H_n^{\rm eff}(a) = \frac{1}{2L}(-id/da)^2 + \frac{\mu^2L}{2} \left(
a - \frac {\pi(2n+1)}{2gL} \right)^2
  \end{eqnarray}
where now \footnote{One and the same letter $\mu$ denotes
different mass terms, depending on the fermion content of the
theory at hand. We hope that this fact will cause no confusion}
 $$
\mu^2 =2g^2/\pi .
$$
 The ground state wave function of the
hamiltonian (\ref{Heff2}) is
  \begin{eqnarray}
  \label{wave2}
|vac_n\rangle = \left(\frac{Lg}{\pi}\right)^{1/4}
{\rm e}^{-\frac{\mu L}2 \left[
a- \frac {\pi(2n+1)}{2gL} \right]^2} \prod_k \psi_L^{(k)}
\prod_k \psi_R^{(k)} \prod_k \chi_L^{(k)}
\prod_k \chi_R^{(k)}
  \end{eqnarray}
where, again, the products involve only the states with the negative
energies.
We can impose now a {\em modified} superselection rule
subdividing the
whole Hilbert space of the hamiltonian (\ref{Ham2}) into the sectors
with
wave functions belonging to a particular irreducible representation
of the
symmetry $\tilde{S}$:
  \begin{eqnarray}
  \label{ssrule2}
\Psi(A^{\tilde{S}}, \psi^{\tilde{S}}, \chi^{\tilde{S}}) =
{\rm e}^{i\tilde{\theta}}
\Psi(A, \psi, \chi)
  \end{eqnarray}
The physical vacuum state is given, as previously, by a superposition
similar to
(\ref{vactet}),
\begin{equation}
\Psi_{\tilde\theta} =
\sum_{n=-\infty}^\infty e^{in\tilde\theta}
|vac_n\rangle\equiv \Psi_1 +\Psi_2
\end{equation}
where
\begin{equation}
\Psi_1 =\sum_{n\,\, odd}{\rm e}^{in\tilde\theta}|vac_n\rangle
,\,\,\,
\Psi_2 =\sum_{n\,\,  even}{\rm e}^{in\tilde\theta}|vac_n\rangle
\end{equation}
are the eigenfunctions of the large gauge transformation $S$.
The standard instanton calculation of any quantity corresponds
to averaging this quantity over $\Psi_1$ or $\Psi_2$, rather than
over the true vacuum state $\Psi_{\tilde\theta}$.
Notice that $\Psi_1$ and  $\Psi_2$ are linear combinations of
$$
\Psi_{\tilde\theta}\,\,\, {\rm and}\,\,\,  \Psi_{{\tilde\theta}'},
\,\,\, {\tilde\theta}' = \tilde\theta +\pi \,\, ({\rm mod}\,\, 2\pi ).
$$

It is instructive to consider the action of the conserved
non-anomalous charges on the vacuum wave function. The vector
charges
$$
\int dx \bar\psi\gamma_0\psi ,\,\,\, {\rm and}
\int dx \bar\chi\gamma_0\chi
$$
can be defined in such a way that the corresponding charge
operators will annihilate $\Psi_{\tilde\theta} $ (i.e. the
charge of the vacuum vanishes). As for the axial charges only one of
them is non-anomalous,
\begin{equation}
Q_5^{(-)} =
\int dx \bar\psi\gamma_0\gamma_5\psi -
\int dx \bar\chi\gamma_0\gamma_5\chi .
\end{equation}
Then it is easy to see that the action of $Q_5^{(-)}$ produces a
different
state,
\begin{equation}
Q_5^{(-)}\Psi_{\tilde\theta} = \Psi_{\tilde\theta} -
\Psi_{{\tilde\theta}'} .
\end{equation}
This means, of course, that $\Psi_{\tilde\theta}\,\,\, {\rm and} \,\,\,
\Psi_{{\tilde\theta}'}$ are degenerate in energy.

The standard vacuum angle $\theta$ is introduced with respect
to the large gauge transformation $S$. Then we see that
the genuine vacuum angle $\tilde{\theta}$ in the model considered
is
twice smaller than the standard angle $\theta$ which enters the
standard
superselection rule (\ref{ssrule}) (with dynamical variables $\chi$
being
included). The angle $\tilde{\theta}$ varies within the interval $(0,
2\pi)$ --
this  means that we {\em must} allow  ${\theta}$ to vary in the
interval
$(0, 4\pi)$. We may conjecture that a similar situation takes place in
the pure Yang-Mills theory and in the
Yang-Mills theory coupled to adjoint fermions where the proper
range for the
vacuum angle $\theta$ is not $(0, 2\pi)$, as often assumed, but
rather $(0, 2\pi N_c)$ (for the unitary gauge groups; for a discussion
of this question see
\cite{LS}).

The fermion condensate $\langle\bar{\psi}_R
\psi_L\rangle_{\tilde{\theta}}$
appears now
due to nonzero matrix elements
$$
\langle vac_1 |\bar{\psi}_R \psi_L|vac_0\rangle , \,\,\,
\langle vac_3 |\bar{\psi}_R
\psi_L|vac_2\rangle ,\,\,\,  {\rm
etc.}
$$
Similarly, the fermion condensate $\langle\bar{\chi}_L
\chi_L\rangle_{\tilde{\theta}}$
appears due to nonzero matrix elements
$$
\langle vac_2 |\bar{\chi}_R \chi_L|vac_1\rangle , \,\,\,\langle vac_4
|\bar{\chi}_R
\chi_L|vac_3\rangle ,\,\,\, {\rm
etc.}
$$
Indeed, it is clear from Fig. 3 that in passing  from $n=0$ to $n=1$
only
the levels of the field $\psi$ cross the zero-energy mark. In passing
from $n=1$ to
$n=2$ only
the levels of the field $\chi$ cross the zero. And then the pattern
repeats
itself.

As a result, we get
  \begin{eqnarray}
   \label{ctet2}
\langle \bar{\psi}_R \psi_L\rangle_{\tilde{\theta}} =
\langle \bar{\chi}_R \chi_L\rangle_{\tilde{\theta}}  = {\rm
e}^{-i\tilde{\theta}} \frac
1{2L}
\exp\{
- \frac {\pi}{2\mu L} \},  \nonumber  \\
\langle\bar{\psi}_L \psi_R\rangle_{\tilde{\theta}} =
\langle\bar{\chi}_L \chi_R\rangle_{\tilde{\theta}}  = {\rm e}^{
i\tilde{\theta}} \frac
1{2L}
\exp\{
- \frac {\pi}{2\mu L} \} .
  \end{eqnarray}

This analysis can be easily generalized to the case of arbitrary
number of
flavors with the boundary conditions (\ref{bc}). The minima of
the effective
potential occur at
  \begin{eqnarray}
  \label{AminN}
a = \frac \pi{NgL} (2n+1),~~~~~~~~n ~~{\rm integer} .
  \end{eqnarray}
The hamiltonian of the system enjoys a new global symmetry
  \begin{eqnarray}
  \label{TN}
\tilde{S_N}:~~~~~~~\left\{
\begin{array}{c}
A_1(x) \rightarrow A_1(x) + \frac {2\pi}{NgL}\\~~~\\
\psi_p(x) \rightarrow {\rm e}^{2\pi i x/(NL)} \psi_{[p+1]_{mod.
N}}(x)\\
\end{array}\right.
  \end{eqnarray}
One can impose the superselection rule which picks out the states
transforming as irreducible representation of $\tilde{S}_N$. This
representation is characterized by a vacuum angle $\tilde{\theta}_N$
varying in the range $(0, 2\pi)$. The ``old" vacuum angle
$\theta = N\tilde{\theta}$ varies in the range $(0, 2\pi N)$.

The fermion condensates are
  \begin{eqnarray}
   \label{ctetN}
\langle\bar{\psi}_R \psi_L\rangle_{\tilde{\theta}_N} =
\langle\bar{\psi}_L \psi_R\rangle_{\tilde{\theta}_N}^* =  {\rm
e}^{-i\tilde{\theta}_N}
 \frac 1{NL} \exp\{
- \frac {\pi}{N\mu L} \}
  \end{eqnarray}
with $\mu^2 = Ng^2/\pi$.

\section{Fermion correlator}

In the previous section, we restricted ourselves to the case $gL \ll
1$
where the Born-Oppenheimer approximation works and the
hamiltonian analysis
is easy. It is interesting, however, to extend our results to arbitrary
$L$,
especially for large $L \gg g^{-1}$ to make contact with the results of
Sect. 2 referring to  the infinite volume. When $gL$ is large,
non-trivial nature of the boundary conditions (\ref{bc}) should be
irrelevant. In particular, the fermion condensate should vanish in the
limit
$L \rightarrow \infty$ as it does in the theory with the standard
boundary conditions.

And, indeed, the condensate vanishes in this limit. In this section, we
calculate  the condensate for arbitrary $L$ employing an
indirect but technically the simplest way. In the next section this
result will be confirmed by a direct computation of the path
integral in the fracton background.

Consider the Euclidean correlator (\ref{ctau}). We show that in
the twisted multiflavor Schwinger model defined on a finite circle,
this correlator tends to a constant in the limit $\tau_0 \rightarrow
\infty$.
The cluster decomposition  implies then the existence of the fermion
condensates.

To begin with, let us describe how the correlator (\ref{ctau}) is
calculated in the standard Schwinger model. For our purposes it is
inconvenient to use the bosonization technique \cite{Jac} which
cannot be
directly generalized to the twisted multiflavor case. We, instead, use
the  functional integral approach developed in \cite{indus,Wipf} and
applied for   calculating  the fermion correlator in \cite{inst}.

Note first of all that the correlator (\ref{ctau}) has the zero chiral
$U_A(1)$ charge and acquires only the contributions from the
topologically
trivial sector. Our calculation is done in two steps. First we calculate
the fermion path integral in a given gauge field background and then
integrate
over the gauge fields. To regularize the path integral in the
infrared let us define the theory on  torus with the spatial
dimension
$L$ and the Euclidean time dimension $T$. The latter will be
assumed very
large and will be sent to $\infty$ as soon as possible.

The path integral for the Schwinger model on  torus has been
carefully
analyzed in Ref. \cite{Wipf}. Any topologically trivial Euclidean field
$A_\mu(x)$ can be represented as
  \begin{eqnarray}
  \label{decomp}
A_\mu(\tau, x) = A_\mu^{(0)} - \epsilon_{\mu\nu}\partial_\nu
\phi(\tau, x)
+ \partial_\mu \lambda (\tau, x)
  \end{eqnarray}
where
$$
A_\mu^{(0)} = \left( \frac {2\pi}{gT} h_0, \frac {2\pi}{gL}
h_1
\right),\,\,\,  0 \leq h_{0,1} \leq 1
$$
is
the constant part of the potential, $
\partial_\mu \lambda (\tau, x)$ is the gauge part, and the part
$- \epsilon_{\mu\nu}\partial_\nu \phi(\tau, x)$ carries nontrivial
dynamical information. Periodic boundary conditions on the
functions
$\phi(\tau, x)$ and $\lambda (\tau, x)$ are imposed in both the
spatial and the
Euclidean time directions. One should also impose the constraint
 \begin{eqnarray}
\label {proj}
\int dx d\tau \phi(\tau, x) = 0 .
 \end{eqnarray}
This constraint excludes the constant part of $\phi$ which is the zero
mode of
the Laplace operator on compact manifolds.
 In two dimensions the fermion determinant
can be explicitly calculated  in {\em any} gauge field background. As
a result, the partition function in the topologically trivial sector can
be written as (including the fermion determinant)
$$
Z_{\nu = 0} = \int  \prod_\mu dA_\mu^{(0)} F(A_\mu^{(0)}) \times
$$
\begin{eqnarray}
  \label{Z0}
\prod_{\tau, x}
d\phi(\tau, x) {\rm exp}\{ - \frac 12
\int_0^Ldx \int_0^Td\tau~ \phi(\Delta^2 - \mu^2 \Delta) \phi \} .
  \end{eqnarray}
A remarkable fact is that the constant harmonics $A_\mu^{(0)}$
completely
decouple here and, in particular, the explicit form of the function
$F(A_\mu^{(0)})$ (which can  be determined, though) is irrelevant.
What {\em is} relevant is that, for very large $T$, the integral is
concentrated
in the region
\begin{equation}
A_0^{(0)} = 0, \,\,\, A_1^{(0)} = \frac \pi{gL}
\label{back}
\end{equation}
 (in this value one
recognizes one of the minima of the effective potential which
happens to lie within
the region of integration $\left(0 \leq A_1^{(0)} \leq \frac
{2\pi}{gL}\right)$
).

The path integral for the fermion correlator (\ref{ctau}) has the form
$$
C(\tau_0)  =
Z_0^{-1} \int  \prod_\mu dA_\mu^{(0)} F(A_\mu^{(0)}) \prod_{\tau,
x}
d\phi(\tau, x) {\rm Tr} \left\{ S_\phi(0,\tau_0) \frac {1-
\gamma^5}2
S_\phi(\tau_0,0) \frac{1+\gamma^5}{2}
\right\}    \times
$$
 \begin{eqnarray}
  \label{cpath}
\exp\{ - \frac 12
\int_0^Ldx \int_0^Td\tau~ \phi(\Delta^2 - \mu^2 \Delta) \phi \}
  \end{eqnarray}
where the spatial coordinates of the initial and final points (set equal
to zero)
are not indicated explicitly. Moreover,
$S_\phi(0, \tau_0)$ is the fermion Green's function in the  given
gauge
field background $-\epsilon_{\mu\nu} \partial_\nu \phi(\tau, x) +
\delta_{\mu 1} \frac \pi{gL}$. Another remarkable simplification
which
occurs in two dimensions is that  Green's function $S_\phi(0,
\tau_0)$
can be found exactly,
  \begin{eqnarray}
  \label{Green}
S_\phi(0, \tau_0) = \exp\{-g\gamma^5 \phi(0)\} S_0(0, \tau_0)
\exp\{-g\gamma^5 \phi(\tau_0))\} ,
  \end{eqnarray}
where $S_0(0, \tau_0)$  is the fermion Green's function estimated
in the  background (\ref{back}).

Substituting Eq. (\ref{Green}) into Eq. (\ref{cpath}), we see that the
path
integrals for the partition function (\ref{Z0}) and for the fermion
correlator (\ref{cpath}) are Gaussian and can be done explicitly (this
is,
of course, a consequence of the fact that the Schwinger model is
exactly
soluble). We get
  \begin{eqnarray}
  \label{CG}
C(\tau_0) = C_0(\tau_0) e^{4g^2[{\cal G}(0) - {\cal G}(\tau_0)]}
  \end{eqnarray}
where $C_0(\tau_0)$ is the fermion correlator in the background
(\ref{back}) and ${\cal G}(x)$ is  Green's function of the operator
${\cal O} = \Delta^2 - \mu^2\Delta$
  \begin{eqnarray}
  \label{eqgreen}
( \Delta^2 - \mu^2\Delta) {\cal G}(x) = \delta^{(2)}(x)
  \end{eqnarray}
Notice that in the limit $T \rightarrow \infty$ torus converts into
cylinder, i.e. a
non-compact manifold, and there is no need to bother with the
elimination of the zero mode of the Laplace operator as is the case
for torus \cite{Wipf} and  sphere \cite{indus}.

The free Euclidean correlator of the fermion densities can be readily
found on $S_1\otimes R_1$,
  $$
C_0(\tau_0) = \frac{1}{2}{\rm Tr} \left \{ \int
\frac{d\omega_1}{2\pi}
{\rm e}^{i\omega_1\tau_0} \frac 1L \sum_{n=-\infty}^\infty
\frac{\omega_1\gamma_0^E + k_n\gamma_1^E} {\omega_1^2 +
k_n^2} \right. \times
$$
\begin{eqnarray}
  \label{Cfree}
\left. \int \frac{d\omega_2}{2\pi}
{\rm e}^{-i\omega_2\tau_0} \frac 1L \sum_{m=-\infty}^\infty
\frac{\omega_2\gamma_0^E + k_m\gamma_1^E} {\omega_2^2 +
k_m^2}
\right\}
  \end{eqnarray}
where $\gamma_0^E = \sigma_2, \gamma_1^E = \sigma_1$ are the
Euclidean $\gamma$
matrices and $k_n = \pi(2n+1)/L$. (The shift $\pi/L$ occurs due
to the
nonvanishing background $A_1^{(0)}$.) Alternatively, we could
impose antiperiodic
boundary
conditions in the spatial direction. In this case the integrals
(\ref{Z0}) and (\ref{cpath}) would pick out the value $A_1^{(0)} = 0$
with the  same final result.

Doing the integrals  through residues (the integration contour is
closed in the upper half-plane for $\omega_1$ and in
the
lower half-plane for $\omega_2$) we arrive at
   \begin{eqnarray}
   \label{Cfreeres}
C_0(\tau_0) = \frac{1}{L^2} \left( \sum_{n=0}^\infty {\rm e}^{-\frac
{\pi\tau_0}
L (2n+1)} \right)^2 = \frac 1{4L^2 \sinh^2 \frac {\pi\tau_0}L } .
   \end{eqnarray}

The next step is calculating  Green's function ${\cal G}(\tau_0)$. We
have
  \begin{eqnarray}
  \label{Greenres}
{\cal G}(\tau_0) = \frac{1}{L} \int_{-\infty}^\infty \frac
{d\omega}{2\pi} {\rm e}^{i\omega \tau_0} \sum_{n=-\infty}^\infty
\frac 1{\lambda_{n\omega} (\lambda_{n\omega} + \mu^2)}
  \end{eqnarray}
where
$$
\lambda_{n\omega} = \omega^2 + \left(\frac {2\pi}L
\right)^2 n^2 .
$$
The sum in Eq. (\ref{Greenres}) can be calculated using the identity
   \begin{eqnarray}
   \label{sum}
\sum_{n=-\infty}^\infty
\frac{1}{n^2+a^2} = \frac{\pi}{a}
\coth \pi a ,
   \end{eqnarray}
see \cite{grad}, Eq. (1.217).
Substituting  Green's function (\ref{Greenres})  obtained in this way
and
$C_0(\tau_0)$ from Eq. (\ref{Cfreeres}) into Eq. (\ref{CG}) we finally
get
$$
C(\tau_0) = \frac 1{4L^2 \sinh^2 \frac {\pi\tau_0}L} \times
$$
\begin{eqnarray}
   \label{Ctaures}
\exp \left\{
\int_{-\infty}^\infty d\omega (1-{\rm e}^{i\omega \tau_0}) \left[
\frac
1\omega
\coth \frac {L\omega }2 - \frac 1{\sqrt{\omega^2 + \mu^2}}
\coth \frac {L\sqrt{\omega^2 + \mu^2}}2 \right]
\right \} .
   \end{eqnarray}
The expression in the exponent has an infrared singularity at
$\omega =0$. It is not difficult to understand how it must be treated.
Recalling that we actually start from torus with a very large $T$ and
eliminate the zero mode of the Laplace operator we conclude that the
integral in the exponent must be understood as the principal value.
Then the factor $(1-{\rm e}^{i\omega \tau_0})$ can be substituted by
$(1-\cos \omega \tau_0)$ and the integral becomes well-defined.

The correlator (\ref{Ctaures}) tends to a constant in the limit $\tau_0
\rightarrow \infty$.
After some algebra it is not difficult to find
$$
C(\tau_0 \rightarrow \infty )
=\left( \frac{\mu}{4\pi}\right)^2 {\rm e}^{(2\gamma - 2I)}
$$
where $\gamma$ is Euler's constant and
\begin{equation}
I =
\int_{0}^\infty \frac {d\omega}
{\sqrt{\omega^2 + \mu^2}} \left( \coth \frac {L\sqrt{\omega^2 +
\mu^2}}2 - 1
\right) .
\end{equation}
The cluster decomposition yields us then the value of
the  fermion condensate, up to a phase factor,
 \begin{eqnarray}
  \label{condres}
\langle\bar{\psi}_R \psi_L\rangle =
\langle\bar{\psi}_L \psi_R\rangle^* = {\rm e}^{-i\alpha} \frac {\mu
{\rm
e}^\gamma}{4\pi}{\rm e}^{-I} ,
  \end{eqnarray}
in agreement with \cite{hosotani,Wipf,cond}. The phase $\alpha$ is
nothing
else than the vacuum angle $\theta$, as follows from the exact Ward
identities (which are
not specific for two dimensions but hold also in the four-dimensional
QCD).

One can be interested not only in the final result but also in the
question of what particular  configuration of the gauge field
(\ref{decomp}) saturates the path integral, or, in other words, what
the  stationary point of the functional integral (\ref{cpath})) is. As
was  shown in  \cite{inst} this is the instanton-antiinstanton
configuration,
  \begin{eqnarray}
  \label{IA}
A_\mu^{stat} (\tau, x) = A_\mu^{inst} (\tau, x) -
A_\mu^{inst} (\tau - \tau_0, x) .
  \end{eqnarray}
The explicit result for $A_\mu^{inst}(\tau, x)$ can be derived for any
$L$
but the results are particularly simple in the limits $gL \gg 1$ and
$gL \ll 1$. As was already mentioned (see \cite{inst} for a detailed
discussion), the question is much more meaningful in the small $gL$
limit
where quantum fluctuations are suppressed and the situation is
quasiclassical. In the gauge $A_0 = 0$, the field $A_1^{inst}$ does not
depend on $x$ and is given by the formula
  \begin{eqnarray}
  \label{inst}
A_1^{inst}(\tau) = \left\{ \begin{array}{c}
\frac \pi{gL}\exp\{\mu\tau\},~~~~~~~~~\tau \leq 0 \\~~\\
\frac \pi{gL}[2-\exp\{-\mu\tau\}],~~~~~~~~~\tau \geq 0 .
\end{array} \right.
  \end{eqnarray}
It has essentially the same form as the fracton configuration in
Eq. (\ref{Afr}), but the photon mass $\mu$ is evaluated at $N=1$
and the
overall factor $1/N$ is absent. The configuration (\ref{inst}) carries
the
unit topological charge (\ref{nu2}). It is, indeed, the instanton.

We described in such details the functional integral calculation
of the
correlator (\ref{ctau}) in the standard Schwinger model because its
generalization to multiflavor case is straightforward.

Consider first the multiflavor model with the standard periodic
boundary
conditions. Again, the correlator is given by the formula (\ref{CG}).
Again,
the functional integrals (\ref{Z0}) and (\ref{cpath}) pick up the
value
$A_1^{(0)} = \frac \pi{gL} $, and the result for the correlator in the
constant
background (\ref{Cfreeres}) is the same as previously.  Green's
function
${\cal G}(\tau_0, 0) $ is  modified a little bit, but the modification is
trivial and amounts to a rescaling of $\mu$. As a result we get
$$
C_N(\tau_0) = \frac 1{4L^2 \sinh^2 \frac {\pi\tau_0}L} \times
$$
\begin{eqnarray}
   \label{CNres}
\exp \left\{ \frac 1N
\int_{-\infty}^\infty d\omega (1-{\rm e}^{i\omega \tau_0}) \left[
\frac
1\omega
\coth \frac {\omega L}2 - \frac 1{\sqrt{\omega^2 + \mu^2}}
\coth \frac {L\sqrt{\omega^2 + \mu^2}}2 \right]
\right \} .
   \end{eqnarray}
The origin of the factor $1/N$ is easy to understand if
turning to Eq. (\ref{CG}) we rewrite $4g^2$
in the exponent as $4\pi \mu^2/N$.

This factor $1/N$ in the exponent has drastic consequences. In
contrast to the correlator (\ref{Ctaures}) $C_N(\tau_0)$ falls off now
exponentially
$$
 C_N(\tau_0 \rightarrow \infty )\propto \exp \left\{ - \frac {2\pi
\tau_0}L \left(
1- \frac
1N \right) \right\}
$$
at large $T$ and the condensate is not generated.

Consider finally the twisted multiflavor case (for simplicity, we
restrict
ourselves to  the case $N=2$). The exponential factor in (\ref{CG}) is
the
same since  Green's function $G(\tau_0)$  knows nothing
about the fermion boundary conditions. What {\em is} modified is
the preexponential
factor $C_0(\tau_0)$ because the constant background in which the
correlator $C_0(\tau_0)$ is evaluated is now different. The
values of $A_1^{(0)}$ on which the path integrals (\ref{Z0}) and
(\ref{cpath}) are
concentrated coincide with the minima of the effective potential
(\ref{Amin2}), i.e. $A_1^{(0)}=\pi (2n+1)/(2gL)$. One can readily
convince oneself that all these minima give  identical
contributions both to the partition function and the correlator (the
latter
property is specific to $N=2$). Calculating the integrals over
$\omega_1$
and $\omega_2$ in Eq. (\ref{Cfree}) we then get
   \begin{eqnarray}
   \label{CNtw}
\tilde{C}^N_0 (\tau_0)= \frac 1{L^2} \left[ \left( \sum_{n=0}^\infty
e^{- \frac {\pi\tau_0}{2L} - \frac {2\pi\tau_0 n}L} \right) ^2
+ \left( \sum_{n=0}^\infty
e^{- \frac {3\pi\tau_0}{2L} - \frac {2\pi\tau_0 n}L} \right) ^2
\right] = \frac {\cosh \frac {\pi\tau_0}L}{2L^2 \sinh \frac
{\pi\tau_0}L} .
   \end{eqnarray}
Substituting it in Eq. (\ref{CG}) with the same exponential as in
Eq. (\ref{CNres}) we see that the correlator does tend now to a
constant in  the
limit $\tau_0 \rightarrow \infty$ (we remind that $N=2$). The
explicit value for the square root of this constant is
\begin{eqnarray}
  \label{condtw}
\langle\bar{\psi}_R \psi_L\rangle  =
\langle\bar{\psi}_L \psi_R\rangle^* =
{\rm e}^{-i\alpha} \sqrt{ \frac {\mu
{\rm e}^\gamma}{16\pi L}}{\rm e}^{-I/2} .
  \end{eqnarray}
The small $gL$ asymptotics of this expression coincides with the
result
(\ref{ctet2}) derived in the previous section. The phase $\alpha$
coincides
with the vacuum angle $\tilde{\theta} = \theta/2$ as  required by
the Ward identities.

At large $gL$ the condensate behaves as
  \begin{eqnarray}
  \label{condLL}
\langle\bar{\psi}_R \psi_L\rangle =
\langle\bar{\psi}_L \psi_R\rangle^* = {\rm e}^{-i\tilde{\theta}}\sqrt{
\frac {\mu
{\rm e}^\gamma}
{16\pi L}}
  \end{eqnarray}
and {\em disappears} in the limit $L \rightarrow \infty$. In is
instructive
to confront the result (\ref{condLL}) with the behavior $\propto
1/\tau_0$
of the correlator (\ref{ctau}) in the infinite volume limit as follows
from
the result (\ref{cinfv}). In the twisted model with finite $L$ this
falloff
is frozen  when $\tau_0$ reaches $L$.  In the  model with the
standard  boundary
conditions the correlator continues to fall off exponentially in the
region $\tau_0 \gg L$.

\section{Fracton path integral}
\setcounter{equation}{0}

This is, perhaps, the central section of the paper. We derive the
result
(\ref{condtw}) {\em directly} by calculating the functional integral
for
the  partition function in the fracton sector with the topological
charge
$\nu = \pm 1/2$ (for simplicity, we restrict ourselves to the
two-flavor case).

In order to do this we must, however, first {\em define} what
this functional integral actually means. The standard definition
  \begin{eqnarray}
  \label{Zwrong}
Z_{\nu = 1/2} = \int \prod_{\tau, x, \mu} dA_\mu(\tau, x)
\det_\psi(i{\not\!\!{\cal D}} - m)
\det_\chi (i{\not\!\!{\cal D}} -m) \exp\left\{- \frac 14 \int d^2x
F_{\mu\nu}^2 \right \}
  \end{eqnarray}
is not suitable here because the Dirac operator for an individual
fermion
$\psi$ or $\chi$ is not defined on a compact manifold with the
background
carrying a fractional topological charge -- self-consistent solutions for
the eigenvalue Dirac equation are absent.

Still, in the case of the twisted boundary conditions (\ref{bc2}) the
configurations with $\nu = \pm 1/2, \pm 3/2, \ldots$ {\em do}
contribute to the partition function.
To understand this we have to go back to basics and recall how
topologically nontrivial path
integrals appear in a field theory in the first place.

 \subsection{Path integral in the standard Schwinger model}

In a simple-minded field theory with no gauge fields and no
topological
effects (say, in the Yukawa theory in four dimensions), the vacuum
partition
function is defined as
$$
 Z = \lim_{T \rightarrow \infty} \sum_n e^{-TE_n} =
\lim_{T \rightarrow \infty} \int \prod_{{\bf x}} d\phi({\bf x})
d\psi(x)
d\psi^+(x)\exp\{-\psi^+(x)\psi(x)\}\times
$$
 \begin{eqnarray}
  \label{Zdef}
{\cal K}[\phi({\bf x}), \psi({\bf x}); ~\phi({\bf x}),
\psi^+({\bf x}); ~T]
  \end{eqnarray}
where
$$
  {\cal K}[\phi({\bf x}), \psi({\bf x}); ~\phi({\bf x}), \psi^+({\bf x}); ~T]
=
$$
\begin{eqnarray}
  \label{K}
\sum_n \Phi_n (\psi({\bf x}), \phi({\bf x})) \left(
\Phi_n (\psi({\bf x}), \phi({\bf x})) \right)^* e^{-TE_n}
  \end{eqnarray}
is the Euclidean evolution operator. The expression (\ref{Zdef}) can
be
written as a path integral
  $$
Z = \lim_{T \rightarrow \infty} \int \prod_{\tau, {\bf x}}
d\phi(\tau,{\bf x})
d\psi(\tau, {\bf x}) d\bar{\psi}(\tau, {\bf x})\times
$$
\begin{eqnarray}
  \label{Zlagr}
\exp \left\{
- \int_0^T d\tau \int d{\bf x} ~{\cal L}[\phi(\tau, {\bf x}),
\bar{\psi}(\tau, {\bf x}), \psi(\tau, {\bf x})]
\right\}
  \end{eqnarray}
with the boundary conditions in the Euclidean time
\footnote{ Then $T$ can be treated as inverse temperature.
The fact that the boundary conditions to be imposed on the fermion
fields
should be antiperiodic in the Euclidean time is widely known,
but its accurate proof involves
some intricacies. A very good pedagogical derivation can be found in
Ref. \cite{Brown}.}
  \begin{eqnarray}
  \label{bcstand}
\phi(\tau + T, {\bf x}) = \phi(\tau, {\bf x}), \nonumber \\
\psi(\tau + T, {\bf x}) = -\psi(\tau, {\bf x}),
  \end{eqnarray}
  (and if the theory is defined on a spatial box some boundary
conditions in spatial directions must be imposed as well).

In gauge theories one should  take into account in the sum
(\ref{Zdef}) only the physical states annihilated by the constraints.
The trace
of the constrained evolution operator can also be written as a path
integral involving additional integrations over the gauge
transformation parameters. Let us  do it for the standard
Schwinger
model defined on  torus,
  $$
Z^{SSM} \stackrel{?}{=} \lim_{T \rightarrow \infty}
\int \prod_{x} d\lambda(x) dA_1( x) d\psi(x) d\psi^+(x)
\exp\{-\psi^+(x)\psi(x)\} \times
$$
\begin{eqnarray}
  \label{Z?}
 {\cal K}[A_1( x), \psi(x); ~A_1^{(\lambda )}({ x}),
\psi^{+(\lambda )}( x); ~T]
  \end{eqnarray}
where $\lambda(x)$ is a periodic gauge transformation function
{\em
continuously deformable to zero} (the integration over such functions
takes
complete account of the constraint (\ref{Gauss}), the Gauss law);
the superscript $(\lambda)$ marks the gauge-transformed
quantities,
the superscript SSM refers to the standard Schwinger model.

It is rather obvious, however, that (\ref{Z?}) is not the full partition
function of the theory but only a part of it corresponding to the
topologically trivial gauge field configurations. Path integrals
calculated
with the prescription (\ref{Z?}) do not satisfy the cluster
decomposition
property. For example, the fermion condensate in the topologically
trivial
sector is zero, but the $\tau_0 \rightarrow \infty$ limit of the
correlator (\ref{ctau}) is not.

To write the physical partition function, one should impose the
superselection rule (\ref{ssrule}) {\em on top} of the Gauss law
constraint
(\ref{Gauss}). The correct partition function in the sector with the
given vacuum angle $\theta$ is given by
  $$
Z^{SSM} = \lim_{T \rightarrow \infty} \sum_n {\rm e}^{-in\theta}
\int \prod_{{x}} d\lambda(x) dA_1({x}) d\psi(x) d\psi^+(x)
\exp\{-\psi^+(x)\psi(x)\} \times
$$
\begin{eqnarray}
  \label{Zsumn}
{\cal K}[A_1({x}), \psi({x}); ~S^n A_1^{(\lambda )}({x}),
S^n \psi^{+(\lambda)}({x}); ~T]
  \end{eqnarray}
where the global symmetry transformation $S$ is defined in
(\ref{T}). Let
us take a closer look, say, at the term with $n=1$. After trading off
the integral over
$\lambda(x
)$ for the integral over $A_0(x)$ (the variable canonically conjugated
to
the Gauss law constraint) and presenting the evolution operator as a
path
integral over the Euclidean time  we get
  \begin{eqnarray}
  \label{Zinst}
 Z_1^{SSM} = {\rm e}^{-i\theta}~\lim_{T \rightarrow \infty}
\int \prod_{\tau, {x}, \mu} dA_\mu({x}) d\bar{\psi}(x) d\psi(x)
\exp\{ -S_E^{SM} (A_\mu, \bar{\psi}, \psi) \}
  \end{eqnarray}
where the integration goes over the fields satisfying the boundary
conditions
$$
A_0(\tau + T, x) = A_0(\tau, x) ,
$$
$$
A_1(\tau + T, x) = A_1(\tau, x) + \frac {2\pi}{gL} ,
$$
\begin{eqnarray}
  \label{bcinst}
\psi(\tau + T, x) = -{\rm e}^{2\pi ix/L} \psi(\tau, x) ,
  \end{eqnarray}
(plus periodicity for all fields  in the spatial direction). The boundary
conditions (\ref{bcinst}) describe the field configurations with the
unit
topological charge, the instantons
  \footnote{We hope that the reader will not be confused
  by a little bit different
  description of the instanton configurations in \cite{Wipf,inst} where
the
  trivial boundary conditions (periodic for bosons and antiperiodic for
fermions) in the
  Euclidean time direction were chosen along with the  nontrivial
boundary conditions in the spatial direction.}.
Likewise, the term $n=2$ in the sum in (\ref{Zsumn}) corresponds to
Euclidean configurations with the double topological charge, etc.

\subsection{Twisted model}

The partition function for the twisted model in the sector with a
given
vacuum angle $\tilde{\theta}$  entering the superselection rule
(\ref{ssrule2}) is given by the sum
  $$
Z = \lim_{T \rightarrow \infty} \sum_n {\rm e}^{-in\tilde{\theta}}
\int \prod_{{x}} d\lambda(x) dA_1({x}) d\psi(x) d\psi^+(x) \times
$$
$$
\exp\{-\psi^+(x)\psi(x)\} d\chi(x) d\chi^+(x)
\exp\{-\chi^+(x)\chi(x)\}\times
$$
\begin{eqnarray}
  \label{Zsumntw}
 {\cal K}[A_1({x}), \psi({x}), \chi(x); ~\tilde{S}^n A_1^{(\lambda )}({x}),
\tilde{S}^n \psi^{+(\lambda )}({x}), \tilde{S}^n \chi^{+(\lambda )}({x});
{}~T] .
  \end{eqnarray}
This expression differs from Eq. (\ref{Zsumn}) by an extra pair of
the fermion
variables and, what is more essential, by the substitution
$\theta \rightarrow \tilde{\theta}, \,\,\, S \rightarrow \tilde{S} $.
Consider again the
term with $n=1$. It can be presented in the path integral form
  $$
 Z_{1/2}^{TSM} = {\rm e}^{-i\tilde{\theta}}~\lim_{T \rightarrow
\infty}
\int \prod_{\tau, {x}, \mu} dA_\mu({x}) d\bar{\psi}(x) d\psi(x)
d\bar{\chi}(x) d\chi(x) \times
$$
\begin{eqnarray}
  \label{Zfract}
\exp\{ -S_E^{TSM} (A_\mu, \bar{\psi}, \psi) \bar{\chi}, \chi) \}
  \end{eqnarray}
where the integration goes over the fields satisfying the boundary
conditions
  $$
A_0(\tau + T, x) = A_0(\tau, x) ,
$$
$$
A_1(\tau + T, x) = A_1(\tau, x) + \frac {\pi}{gL} ,
$$
$$
\psi(\tau + T, x) = -{\rm e}^{\pi ix/L} \chi(\tau, x) ,
$$
\begin{eqnarray}
  \label{bcfract}
\chi(\tau + T, x) = -{\rm e}^{\pi ix/L} \psi(\tau, x) ,
  \end{eqnarray}
and the superscript TSM refers to the twisted Schwinger model.
The shift in $A_1$ is twice  smaller compared to the instanton case
(\ref
{bcinst}), and the topological charge of the gauge field is also twice
smaller. It is a fracton field configuration (thereby the subscript
$1/2$  in Eq. (\ref{Zfract})).

We understand now how the difficulty of defining the Dirac operator
in the fracton background is resolved. We see that here the two
fermion fields are
coupled to each other by the boundary conditions and, though we
cannot
formulate the eigenvalue problem for an individual fermion, the
coupled-channels eigenvalue problem
  \begin{eqnarray}
  \label{eigen}
\left\{ \begin{array}{c} {\not\!\!{\cal D} }\psi_n = \mu_n \psi_n ,\\
\\
{\not\!\!{\cal D}} \chi_n = \mu_n \chi_n \end{array}  \right.
  \end{eqnarray}
with eigenfunctions $\psi_n(\tau, x), \,\,\, \chi_n(\tau, x)$ satisfying
the
boundary conditions (\ref{bcfract}) is perfectly well-defined. Hence,
the path  integral (\ref{Zfract}) is well-defined too. The only
remaining problem is to calculate it.

We are interested, eventually, in   the fermion condensate
$\langle\bar{\psi}_R \psi_L \rangle$ which in the limit of very small
mass is given by the ratio
   \begin{eqnarray}
   \label{frZ}
\langle\bar{\psi}_R \psi_L \rangle = - \lim_{m \rightarrow 0} \frac
{Z_{1/2}}{mZ_0} .
   \end{eqnarray}
It is worth reminding that the condensate is the logarithmic
derivative with respect to the mass of the full
partition
function as in Eq. (\ref{cvar}), but the higher-$n$ terms in the sum
(\ref{Zsumntw}) involve the suppression factor $m^{|n|}$ and do not
contribute in the limit $m \rightarrow 0$.

Let us first carefully calculate
 the partition function in the topologically trivial sector
following  the technique developed in Ref. \cite{Wipf}. In the sector
$\nu = 0$ there is no problem whatsoever with calculating the
integrals over
$d\bar{\psi}
d\psi$ and over $d\bar{\chi} d\chi$ separately, and we get
  $$
Z_0^{TSM} = {\cal N} \int_0^1 \int_0^1 dh_0 dh_1\times
$$
\begin{eqnarray}
  \label{Z0tw}
 \int \prod_{\tau, x}
d\phi(\tau, x)
\exp\left\{ - \frac 12 \int d^2x \phi \Delta^2 \phi \right \}
\det_{\psi} (i {\not\!\!{\cal D}}) \det_{\chi} (i{\not\!\!{\cal D}})
  \end{eqnarray}
where $h_{0,1}$ and $\phi(\tau, x)$ are the gauge invariant
variables
characterizing the field $A_\mu(\tau, x)$ as written in
Eq. (\ref{decomp})
and ${\cal N}$ is an unspecified normalization factor. Notice that the
mass term can be set equal to zero here. The
determinants of
the Dirac operator on  torus involve an intricate dependence on the
constant harmonics $h_0, h_1$ through the Jacobi $\Theta$
function; in the limit of
very large $T$ the results are greatly simplified, however. We have
  $$
\det_\psi(i{\not\!\!{\cal D}}) \approx {\rm e}^{\pi T/6L} \exp \left\{
- \frac {2\pi
T}L
\left(h_1 - \frac 12 \right)^2 \right\} \exp \left \{ \frac {g^2}{2\pi}
\int d^2x \phi \Delta \phi \right \},
$$
$$
\det_\chi(i{\not\!\!{\cal D}}) \approx {\rm e}^{\pi T/6L} \left[ \exp
\left\{ - \frac
{2\pi T}L
h_1^2 \right\} + \exp \left\{ - \frac {2\pi T}L
\left(h_1 - 1 \right)^2 \right\} \right] \times
$$
\begin{eqnarray}
  \label {Det0}
\exp \left \{ \frac {g^2}{2\pi} \int d^2x \phi \Delta \phi \right \}
  \end{eqnarray}
($T \gg L, \,\,\, 0 \leq h_1 \leq 1$). Substituting (\ref{Det0}) in
(\ref{Z0tw})
and integrating over $h_1$, we get
  \begin{eqnarray}
  \label{Z0res}
Z_0^{TSM} = \frac {{\cal N}}{\sqrt{T/L}} {\rm e}^{\pi T/12L}
{}~ \int \prod_{\tau, x} d\phi(\tau, x)
\exp\left\{ - \frac 12 \int d^2x \phi (\Delta^2 - \mu_2^2 \Delta)\phi
\right \} .
  \end{eqnarray}

Now, let us find the fracton contribution to the partition function. The
result of integration over the fermion variables in (\ref{Zfract}) is the
determinant of the {\em matrix} Dirac operator (\ref{eigen}) (i.e. the
product of all its eigenvalues $\mu_n$). Fortunately, we need not
calculate
it anew but may use again the results of Ref. \cite{Wipf} exploiting
the following   trick. Consider the function
   \begin{eqnarray}
   \label{PSI}
\Psi(\tau, x) = \psi(\tau, x) + \chi(\tau, x) .
   \end{eqnarray}
This function  does not satisfy definite boundary conditions on the
original torus
$(T,
L)$ but it does satisfy the periodic boundary conditions on the twice
``thicker" torus $(T, 2L)$
$$
 \Psi(\tau, x+2L) = \Psi(\tau, x)
$$
(we will call it ``large" torus).
The fields $\psi(\tau, x)$ and $\chi(\tau, x)$ may be considered as
the sum
of all even and, correspondingly, all odd Fourier spatial harmonics of
$\Psi(\tau, x)$ defined on the box $(T, 2L)$.

Consider now the Dirac operator for the field $\Psi$ on the large
torus
in the gauge field background which consists of two copies: the field
$A_\mu(\tau, x)$ in the interval $L \leq x \leq 2L$ coincides
identically
with the field in the interval $0 \leq x \leq L$.

It is not difficult to see now that the eigenvalues of the Dirac
operator
thus defined coincide identically with the eigenvalues of the original
matrix Dirac operator in Eq. (\ref{eigen}) [the simplest way to
establish
the equivalence of the two fermion path integrals is to write them in
terms
of Fourier harmonics of the fields $\psi(\tau, x)$,  $\chi(\tau, x)$, and
$\Psi(\tau, x)$].

The field $A_\mu(\tau, x)$ defined on the large torus carries the
topological charge $2 \times \frac 12  = 1$. Thus, the fermion path
integral in (\ref{Zfract}) coincides with the determinant of the Dirac
operator in the instanton background. The latter has been calculated
in
\cite{Wipf}. The result is
   \begin{eqnarray}
   \label{DetPsi}
\det_\Psi (i{\not\!\!{\cal D}} - m) = m \sqrt{LT} ~\frac 1{2LT} \int
d^2x {\rm
e}^{-2g\phi(\tau , x)} \exp \left \{ \frac {g^2}{2\pi} \int d^2x \phi
\Delta \phi \right \}
   \end{eqnarray}
where the limits of integration are $0 \leq \tau \leq T, \,\,\, 0 \leq x
\leq
2L$
and the field $\phi$ satisfies the condition $\phi(\tau, x+L) =
\phi(\tau,
x)$.

The factor $m$ in Eq. (\ref{DetPsi}) comes from one complex
zero mode of the
field
$\Psi$ in the  instanton background (and, correspondingly, one zero
mode for
the fields $\psi(\tau , x)$,  $\chi(\tau , x)$ in a fracton background).
The
product of all nonzero eigenmodes is not sensitive to $m$ if it is
small
enough and just coincides with Eq. (3.9) of Ref. \cite{Wipf}.
Substituting
the
result (\ref{DetPsi}) in the path integral for $Z_{1/2}$ and
substituting
\[ \frac 1{2LT} \int d^2x{\rm
e}^{-2g\phi(\tau, x)} \rightarrow{\rm e}^{-2g\phi(0)} \]
(we can do this due to the translational invariance of the path
integral), we
get
  \begin{eqnarray}
  \label{Zfrres}
Z_{1/2} = {\rm e}^{-i\tilde{\theta}} {\cal N}m {\sqrt{TL}}
\int \prod_{\tau, x} d\phi(\tau, x) {\rm e}^{-2g\phi(0)}
\exp\left\{ - \frac 12 \int d^2x \phi (\Delta^2 - \mu_2^2 \Delta)\phi
\right \} .
  \end{eqnarray}
Here the integral runs over the {\em original} torus and the factor 2
in
(\ref{DetPsi}) is taken into account in the definition of $\mu_2^2 =
2g^2/\pi$.
The normalization factor ${\cal N}$ is the same as in Eq. (\ref{Z0res}).
Substituting Eqs. (\ref{Zfrres}) and (\ref{Z0res}) in Eq. (\ref{frZ})
we arrive at the following expression for the fracton contribution
in the fermion condensate:
  \begin{eqnarray}
  \label{condG}
\langle\bar{\psi}_R \psi_L\rangle  = - {\rm e}^{-i\tilde{\theta}}
\frac 1L {\rm e}^{- \pi T/12L}
{\rm e}^{2g^2 {\cal
G}(0)}
  \end{eqnarray}
where ${\cal G}$ is the Green's function of the operator
$\Delta^2 - \mu_2^2 \Delta$ defined in Eq. (\ref{Greenres}) with
$\tau_0$ set equal to zero.
Performing the sums by means  of Eq. (\ref{sum}) we reproduce the
result  (\ref{condtw}) obtained in the previous section.

\section{Conclusions}

It is known for a long time that in the standard Schwinger model on
a circle the topology of the functional space is nontrivial. The
functional space of the gauge fields contains a non-contractible path,
$S_1$. Since the configurational space is also $S_1$ the path integral
is decomposed in a sum corresponding to different winding numbers
in the mapping $S_1\rightarrow S_1$. The size of the
non-contractible path (i.e. what points are to be identified) is
determined by the large gauge transformation.

In this work we suggested a model (the twisted Schwinger model on
a circle) which exhibits a remarkable feature. The non-contractible
path is still there, but the identification of points in the functional
space is decided by
a transformation of fields that is {\em not} pure gauge. Thus, in the
two-flavor twisted model the original circle of the standard
Schwinger model is further glued in such a way that two points lying
on one diameter are identified. It seems quite plausible that this
lesson
may be of a paramount importance for the non-abelian gauge
theories, in solving such long-standing problems as, say, the problem
of Witten's index in $O(N)$ SUSY Yang-Mills theories.

\section{Acknowledgements}
We are indebted to I. Kogan, N. Nekrasov, A. Vainshtein, and A.
Zhitnitsky for illuminating discussions. A. S. acknowledges  warm
hospitality extended to him during his stay at TPI, University of
Minnesota.

This work was supported in part by DOE under the
grant number DE-FG02-94ER40823.

\newpage

{\bf Figure captions}

{\bf Fig. 1}. Fracton-antifracton configuration.

{\bf Fig. 2}. Fermion levels in the standard Schwinger model.

{\bf Fig. 3}. Fermion levels in the twisted two-flavor Schwinger
model.

\end{document}